\documentclass[useAMS,usenatbib]{mn2e}

\usepackage{epsfig,epstopdf,color}

\def\gtrsim{\lower 2pt \hbox{$\, \buildrel {\scriptstyle >}\over
{\scriptstyle \sim}\,$}}
\def\lesssim{\lower 2pt \hbox{$\, \buildrel {\scriptstyle <}\over
{\scriptstyle \sim}\,$}}

\def\suzaku{{\sl Suzaku}}

\def\suzaku{{\sl Suzaku}}

\def\ovii{O~{\scriptsize VII}}
\def\oviii{O~{\scriptsize VIII}}

\def\ka{K$\alpha$}
\def\xs{\ovii~K$\alpha$}

\title[X-ray Emission from CX in the Cygnus Loop SNR]{X-ray Emission from Charge Exchange in the Cygnus Loop SNR}
\author[Roberts \& Wang]{Shawn R. Roberts$^{1}$\thanks{E-mail: srrobert@astro.umass.edu}, Q. Daniel Wang$^{1}$\\
$^{1}$Department of Astronomy, University of Massachusetts, Amherst, MA 01002, USA\vspace{-0.5cm}}
\date{Submitted to MNRAS, September, 2014\vspace{-0.5cm}}
\begin{document}

\pagerange{\pageref{firstpage}--\pageref{lastpage}} \pubyear{2014}

\maketitle

\label{firstpage}

\begin{abstract}
The Cygnus Loop has been the focus of substantial debate concerning the contribution of charge exchange (CX) to supernova remnant (SNR) X-ray emission. We take advantage of a distinct feature of CX, enhanced K$\alpha$ forbidden line emission, and employ the energy centroid of the OVII K$\alpha$ triplet as a diagnostic.  Based on X-ray spectra extracted from an extensive set of Suzaku observations, we measure the energy centroid shifts of the triplet on and off the shock rim of the remnant.  We find that enhanced forbidden to resonance line emission exists throughout much of the rim and this enhancement azimuthally correlates with non-radiative H$\alpha$ filaments, a tracer of strong neutral-plasma interaction in the optical.  We also show that alternative mechanisms cannot explain the enhancement observed.  These results demonstrate the need to model the CX contribution to the X-ray emission of SNRs, particularly for shocks propagating in a partially neutral medium.  Such modelling may be critically important to the correct measurements of the ionization, thermal, and chemical properties of SNRs.
\end{abstract}

\begin{keywords}
atomic processes; ISM: supernova remnants; ISM: abundances; ISM: individual: Cygnus Loop; X-rays: ISM
\vspace{-0.5cm}
\end{keywords}

\section{Introduction}
\label{s:intro}

X-ray astrophysics is layered with a host of interesting phenomena.  As spectral capabilities continue to grow within the X-ray regime, our ability to discern the importance of these varied phenomena has led to many interesting surprises.  One X-ray emission mechanism whose importance we are increasingly becoming aware of is charge exchange or charge transfer (CX; e.g., Lallement 2004).  While CX reactions have been studied in an astrophysical context for decades, their importance for X-ray astrophysics wasn't realized until comet observations in 1996 \citep{Lisse1996,Cravens1997}.  Since this cometary epiphany, CX has been proposed and studied in a wide range of X-ray sources, from planetary atmospheres \citep{Dennerl2002,Dennerl2006,Kharchenko2008} through individual star-forming complexes \citep{Townsley2011} and star-forming galaxies \citep{Tsuru2007,Liu2011,Liu2012} to the cores of clusters \citep{Fabian2011}.  More recently, the first evidence for X-ray emitting CX in a supernova remnant (SNR) has been suggested \citep{Katsuda2011}.

CX is a special case of inelastic collision in which one or more electrons is exchanged between an ion and typically a neutral atom, most likely hydrogen.  As the colliding particles approach each other, their potentials overlap, allowing the photonless exchange and subsequent cascade of the captured electron.  This is an extremely efficient process with a relatively large reaction cross section (about 5 orders of magnitude greater than for electron collisional excitation).  If the capturing ion is sufficiently ionized (e.g. OVIII) then the cascading is expected to result in X-ray emission.  Thus, in principle, X-ray emitting CX can be an important process at any astrophysical site where heavily ionized species interact with neutrals, making the thin post-shock region of SNRs a very promising site for CX.

While CX has been used to explain the broad component of H$\alpha$ emission in many SNRs for more than 30 years \citep{Chevalier1980,Ghavamian2001}, X-ray emission produced by CX has not yet been directly apparent.  This is at least partly due to the lack of a non-dispersive high-resolution spectrometer.  We are instead confined to X-ray CCD data for our studies of diffuse emission.  At the spectral resolution of such data, the pure line emission of CX can mimic the continuum + line emission of a thermal plasma, making it difficult to disentangle the relative contributions of the different emission mechanisms.  Thus, without accounting for the contribution from CX, the modelling of the properties of the shocked plasma (e.g. chemical abundances) can potentially be misleading.  

To delve into the importance of CX for SNR emission we need to be able to spatially and spectrally resolve the nuances of the plasma evolution in nearby remnants, such as the Cygnus Loop SNR.  This remnant has an angular diameter of $\approx$ 3$^{\circ}$ and suffers relatively little absorption.  Such a close SNR provides a unique laboratory for the study of shock physics and non-equilibrium conditions.  These spatial and spectral studies have been done to some extent in the X-ray regime with Suzaku and XMM-Newton, for which observations cover nearly the entire remnant.  However, these studies have seemingly created more questions than answers.

\begin{figure}
\centerline{\epsfig{figure=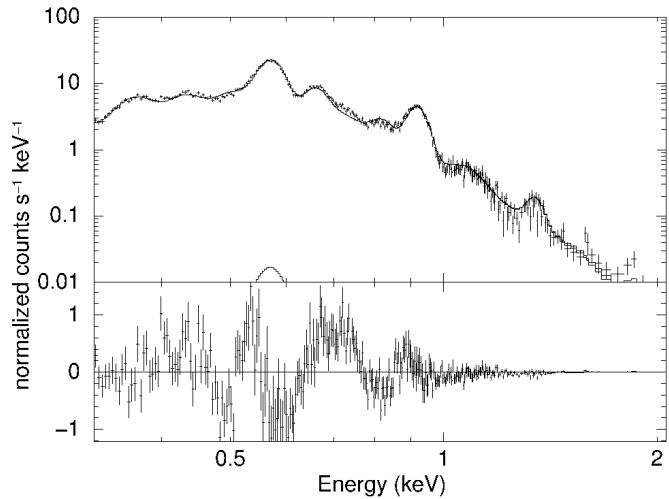,width=0.5\textwidth,angle=0}}
\caption{%\footnotesize 
Typical spectrum of the immediate post-shock region, taken from the NE4 observation (see Section ~\ref{s:obs} for more details).  Spectrum is fit with an absorbed single temperature NEI plasma with best fit abundance parameters listed in Table ~\ref{t:spec_pars}.}
\label{fig:P18_specs}
\end{figure}

With Suzaku, spectral fitting of the immediate post-shock region has led to a quagmire of bewilderment where derived metal abundances appear to fall into two general categories \citep{Uchida2009}.  In some of the rim regions, abundances have been cited as 0.5-1 times the solar value, which is consistent with the surrounding ISM \citep{Cartledge2006,Nieva2012}.  Throughout the rest of the rim, abundances are found to be about 0.2 times the solar value or less, well outside the dispersion of local ISM measurements (e.g. Table ~\ref{t:spec_pars} and Figure ~\ref{fig:P18_specs} for the results of a typical spectral fitting in the immediate post-shock region).  This peculiar situation has led to the consideration of phenomena that are either generally thought to be unimportant or more exotic and poorly understood, with no clear answer emerging.

\begin{table}
 \centering
 \label{t:spec_pars}

  \begin{tabular}{| c | c | }
C & 0.164 ($\pm$0.016) \\  
N & 0.130 ($\pm$0.020) \\  
O & 0.105 ($\pm$0.009)\\  
Ne(=Mg) & 0.230 ($\pm$ 0.016) \\  
Si(=S) & 0.243 ($\pm$0.043) \\  
Fe (=Ni) & 0.125 ($\pm$0.011) \\  
\end{tabular}
  \caption{Typical best fit metal abundances (fraction of solar) for the immediate post shock region, taken from the NE4 observation (see Section ~\ref{s:obs} for more details).}
\end{table}

Missing physics in the modelling of moderate resolution spectra is probably the most natural and intuitive explanation for the regions of apparent low abundance.  \cite{Miyata2008} suggested that resonance-line scattering might account for the regions of relatively suppressed abundances.  However, the authors found that even with assumptions that are likely to amplify the effects of scattering, it could not account for the level of abundance suppression found, particularly that of oxygen.  Reasoning along the same lines, \cite{Katsuda2008} and \cite{Tsunemi2009} posited that the continuum could be higher in these regions due to a nonthermal contribution, thus creating the appearance of lower metal abundances.  Although this did improve the quality of the fits and increased the abundances, the predicted radio fluxes are orders of magnitude higher than the observed values.

With strong line emission, CX could significantly affect the abundance measurement.  \cite{Katsuda2011} noticed when excluding energies around 0.7 keV, the abundances typically drop to those of the low abundance regions.  They posited that the high abundances may be artificial, probably due to CX contributions at $\sim$ 0.7 keV from higher order transitions of \ovii\ (note the excess emission around this energy in Figure ~\ref{fig:P18_specs}).  Also, this enhancement is found to anti-correlate with non-thermal radio emission.  This led the authors to argue that the cosmic-ray precursor more effectively ionizes the pre-shock ISM, thus eliminating the CX process.  However, if this interpretation were to prove valid, it would also necessitate an explanation for the overall low abundances throughout the rim.  Furthermore, the rim regions where the enhancement occurs do not correlate with non-radiative H$\alpha$ emission (an indicator of neutral gas in the post-shock plasma), as would be expected for CX.

Here we propose to judge the importance of CX based on a less model-dependent diagnostic.  In this work we will exclusively use the \xs\ triplet ($\sim0.57$ keV), which is often the strongest line in the spectra.  The triplet consists of the resonance (574.0 eV), inter-recombination (568.6 eV) and forbidden (561.1 eV) transitions.  Effects of CX become clear when the so-called G ratio [$=(f +i)/r$, forbidden$+$inter-combination to resonance transitions] of the triplet is analyzed \citep{Liu2011,Liu2012}.  Due to the cascading and high number of quantum states leading to the forbidden line, the ratio is calculated to be enhanced relative to thermal emission (due to collisional excitation).  Even if unable to resolve the individual components of the triplet, one may adopt its central energy as a proxy for the G ratio, which can typically be determined to an accuracy of more than 20 times better than the intrinsic spectral resolution when the signal to noise ratio is sufficiently high, which is typically the case here \citep{Wargelin2008}.  Therefore, the centroid analysis can be useful, even with a spectrum of moderate resolution.  For the \xs\ triplet, the centroid is expected to be shifted to lower energies by 4.3$(\pm0.5)$ eV for CX emission compared to thermal \citep{Snowden2009}.  This enhancement in the forbidden line is illustrated in Figure ~\ref{fig:P18_specs} by the large residuals on the low energy side of the \xs\ line.

Our observations and data analysis are described in \S\ ~\ref{s:obs}.  We explore how the central energy of the triplet changes azimuthally and with distance from the shock front in \S\ ~\ref{s:spec}.  The spatial distribution of the energy shift is then compared directly to a tracer of strong plasma-neutral interaction in \S\ ~\ref{s:ha}.  Lastly, we compare these results with previous studies of the Cygnus Loop SNR and discuss the implications and alternative scenarios in \S\ ~\ref{s:dis}.
 
\section{Observations and Data Reduction}\label{s:obs}

\begin{figure}
\centerline{\epsfig{figure=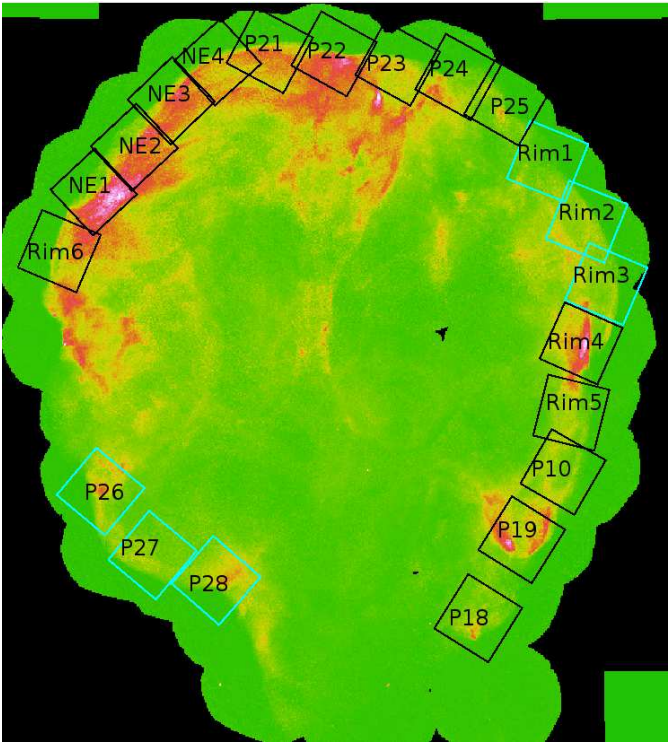,width=0.5\textwidth,angle=0}}
\caption{%\footnotesize 
Cygnus Loop SNR regions observed with \suzaku\ (17.8$'$ x 17.8$'$) overlaid on the logarithmically scaled \textit{Rosat} HRI image.  Cyan observation regions do not show an energy shift of the \xs\ line from the rim to the inner region while black ones do (see text for details).}
\label{f:cygnus_full}
\end{figure}

\suzaku , the ideal telescope for this study, was launched on July 10, 2005 carrying 4 CCDs with independent X-ray telescopes sensitive to photons in the 0.2-12 keV  range \citep{Kunieda2006}.  While three of the CCDs are front-illuminated, one (XIS1) is back-illuminated and is more sensitive to soft X-rays.
Thus, the XIS1 spectrometer (with a spectral resolution of $\sim$ 60 eV at the energy of the OVII K$\alpha$ triplet) is used for the present study.  Observations of the Cygnus Loop with \suzaku\ encompass nearly the entire rim of the SNR, excluding the region of the southern blowout and a small area in the eastern face of the shock front (Table ~\ref{t:obs_snr}; Figure ~\ref{f:cygnus_full}).

\begin{table*}
 \centering
 \begin{minipage}{140mm}\label{t:obs_snr}
  \caption{\suzaku\ XIS1 Observations of the Cygnus Loop SNR}
  \begin{tabular}{| l | l | l | l | l | c | c |}
  \hline\hline
% GC used in analysis
Obs ID & l & b & Start & End & $t_o$  & $t_{e}$\\ \hline
500020010 (NE1) & 75.61 & -8.84 & 2005-11-23 17:39:01 & 2005-11-24 04:55:24 & 20.34 & 14.36 \\
500021010 (NE2) & 75.66 & -8.56 & 2005-11-24 04:56:05 & 2005-11-24 16:14:24 & 21.41 & 20.88 \\
500022010 (NE3) & 75.72 & -8.28 & 2005-11-29 17:47:47 & 2005-11-30 05:39:09 & 21.13 & 17.34 \\
500023010 (NE4) & 75.72 & -7.99 & 2005-11-30 05:41:02 & 2005-11-30 18:23:14 & 25.00 & 19.02 \\
501020010 (P10) & 75.13 & -7.94 & 2007-11-13 06:01:00 & 2007-11-13 14:30:15 & 16.80 & 10.19 \\
501035010 (P18) & 72.85 & -8.69 & 2006-12-18 01:16:09 & 2006-12-18 08:10:21 & 12.01 & 8.71 \\
501036010 (P19) & 73.01 & -8.29 & 2006-12-18 08:11:18 & 2006-12-18 19:47:24 & 18.60 & 17.92 \\
503057010 (P21) & 75.60 & -7.76 & 2008-06-02 22:33:53 & 2008-06-03 07:08:24 & 16.17 & 10.24 \\
503058010 (P22) & 75.39 & -7.53 & 2008-06-03 07:09:17 & 2008-06-03 18:03:14 & 19.29 & 16.49 \\
503059010 (P23) & 75.17 & -7.32 & 2008-06-03 18:04:23 & 2008-06-04 03:32:14 & 19.46 & 15.46 \\
503060010 (P24) & 74.93 & -7.14 & 2008-06-04 03:33:03 & 2008-06-04 15:03:09 & 18.50 & 13.50 \\
503061010 (P25) & 74.69 & -7.04 & 2008-06-04 15:04:06 & 2008-06-05 03:34:24 & 26.00 & 21.16 \\
503062010 (P26) & 74.46 & -9.71 & 2008-05-13 02:53:19 & 2008-05-13 13:08:19 & 16.92 & 14.62 \\
503063010 (P27) & 74.06 & -9.70 & 2008-05-13 13:09:08 & 2008-05-14 01:11:14 & 22.78 & 16.94 \\
503064010 (P28) & 73.77 & -9.54 & 2008-05-14 01:12:11 & 2008-05-14 12:48:14 & 18.17 & 15.72 \\
504005010 (Rim1) & 74.35 & -7.07 & 2009-11-17 22:40:59 & 2009-11-18 22:46:24 & 40.75 & 27.30 \\
504006010 (Rim2) & 74.01 & -7.10 & 2009-11-18 22:47:37 & 2009-11-19 11:37:24 & 26.31 & 14.89 \\
504007010 (Rim3) & 73.12 & -7.21 & 2009-11-19 11:38:17 & 2009-11-20 02:53:24 & 21.56 & 17.63 \\
504008010 (Rim4) & 73.57 & -7.49 & 2009-11-20 02:54:21 & 2009-11-20 08:35:24 & 12.10 & 5.63 \\
504009010 (Rim5) & 73.33 & -7.73 & 2009-11-20 08:40:05 & 2009-11-20 19:05:17 & 15.85 & 12.05 \\
504010010 (Rim6) & 75.49 & -9.15 & 2009-11-20 19:10:13 & 2009-11-21 04:11:24 & 14.33 & 10.44 \\
\hline
\end{tabular}
\end{minipage}
\end{table*}
 
The observations are reduced according to the standard procedures described in the \suzaku\ ABC guide, using HEAsoft version 6.11.1 and CALDB version 2011-11-09.  The data is reprocessed using the ftool \textit{aepipeline}.  When available, data from the 5x5 and 3x3 editing modes are combined.  Two selections are placed on the data, cut-off rigidity (COR)$>$8 and elevation from Earth (ELV)$>$10$^{\circ}$.  The COR selection is used to limit the contamination from the non-X-ray background (discussed below).  The elevation selection is made to keep the Oxygen 0.54 keV flourescent line to a minimum \citep{Smith2007}, which would be blended with the OVII K$\alpha$ triplet, potentially leading to problematic results. The channels of all spectra are adaptively grouped to have a minimum (background-subtracted) S/N threshold of 3 per bin before fitting.

Since our study is based on the differential line centroid of the OVII K$\alpha$ triplet on and off the SNR shock, it is important to check whether or not any systematic instrumental gain shift may affect our measurements.  Investigation of the CCD shows no apparent systematic shift along the CCD.  Thus, we can compare the relative line centroids between regions within an observation without having to consider the actual gain error of the instrument.

\section{Spectral Analysis}\label{s:spec}

Our spectral analysis is conducted with the XSPEC software package (version 12.7.1).  In order to characterize the shock region of the Cygnus Loop SNR, spectra are extracted from 2' wide cuts along the shock front.  Following the findings of \citep{Katsuda2011}, in which the authors determined any enhancements at $\sim$ 0.7 keV occurred within 4' of the shock front, spectra were extracted from the innermost regions of the same observations up to 4' from the shock front in order to characterize the inner regions of the SNR.  Such a spatial distribution is also consistent with previous theoretical work \citep{Lallement2004}.  These two spectra are then compared directly to each other.

The background can arise from three general sources: extragalactic, particle, and locally (SXB).  The particle background is modelled using night Earth observations with the ftool \textit{xisnxbgen} and is subtracted from the source spectra in XSPEC before spectral fitting.  The SXB contribution due to the Local Bubble and solar wind CX at Earth's outer atmosphere and the heliosphere, is modelled as an unabsorbed single temperature thermal plasma, which has been found to give adequate fits despite the contested importance of CX to the SXB emission \citep{Anderson2010,Smith2014}.  The temperature of this local plasma component is fixed to the fiducial value previously found, 0.11 keV \citep{Anderson2010}.  This local emission is dwarfed by the SNR emission and contributes only about 1\% of the emission in soft X-rays.  The extragalactic component, which dominates the emission at high energies after particle background removal, is modelled as a power law with a photon index of $\Gamma$=1.4 \citep{Yoshino2009} together with a foreground absorption column density of 3 $\times$ 10$^{20}$ cm$^{-2}$ \citep{Kosugi2010}.

In order to characterize the center of the \xs\ line, it must be modelled as a separate gaussian.  First, we modelled the source emission as a non-equilibrium ionization (NEI) plasma with floating abundances of each element individually and foreground absorption column density frozen at 3 $\times$ 10$^{20}$ cm$^{-2}$.  After this initial fit is obtained, the oxygen abundance is set to 0 and all other parameters are frozen.  The Oxygen emission is then modelled as a set of lines:  OVII K$\alpha$, OVIII Ly$\alpha$, OVII K$\beta$, OVII K$\gamma$, OVII K$\delta$, OVII K$\epsilon$, OVIII Ly$\beta$, OVIII Ly$\gamma$, OVIII Ly$\delta$, and OVIII Ly$\epsilon$ (all line energies fixed to theoretical values except for the \ovii\ \ka\ and \oviii\ Ly$\alpha$ lines) as done in \cite{Katsuda2011}.  The lines are not expected to be resolved and are set to have a width of 0 keV.

The difference in the OVII K$\alpha$ centroid energy between the inner and rim spectra is presented in Table ~\ref{t:dir_anal} for each observation.  While several of the observations are consistent with no shift, most of the observations do exhibit an apparent shift as expected for CX emission.  Comparing the observations that have an apparent centroid shift (i.e., $\gtrsim$ 3 eV) to Figure ~\ref{f:cygnus_full}, we see that nearly the entire rim of the SNR sees a shift to higher energies from the rim inward with the exception of two regions, one in the northwest rim and another in the southeast rim.  These two regions also have the lowest downstream X-ray flux, with the exception of the blowout, due to the relatively lower density of the ISM in these regions.  The lack of a centroid shift in these regions is probably a natural consequence of a shock propagating into a low-density medium, which also presumably has a relatively small neutral hydrogen fraction and hence little CX emission is expected.

%\begin{table*}
\begin{table}
 \centering
 %\begin{minipage}{140mm}
 \label{t:dir_anal}
  \caption{OVII K$\alpha$ centroid line difference between rim and inner spectra ($E_{inner}-E_{rim}$).  Errors are at the 1$\sigma$ confidence level.}
  \begin{tabular}{| c | c }%| c | c |}
  \hline\hline
Obs ID & $\Delta$E (eV) \\ \hline
501020010 (P10) & 6.0 ($\pm$ 1.41) \\
501035010 (P18) & 4.9 ($\pm$ 1.35) \\
501036010 (P19) & 3.1 ($\pm$ 0.67) \\
503057010 (P21) & 5.1 ($\pm$ 0.86) \\
503058010 (P22) & 3.8 ($\pm$ 0.58) \\
503059010 (P23) & 3.7 ($\pm$ 0.82) \\
503060010 (P24) & 3.9 ($\pm$ 1.05) \\
503061010 (P25) & 4.0 ($\pm$ 1.26) \\
503062010 (P26) & 0.9 ($\pm$ 0.92) \\
503063010 (P27) & -1.0 ($\pm$ 0.95) \\
503064010 (P28) & -2.0 ($\pm$ 0.79) \\
504005010 (Rim1) & -0.9 ($\pm$ 1.39) \\
504006010 (Rim2) & -0.9 ($\pm$ 1.04) \\
504007010 (Rim3) & 0.0 ($\pm$ 0.93) \\
504008010 (Rim4) & 5.0 ($\pm$ 1.41) \\
504009010 (Rim5) & 8.0 ($\pm$ 1.83) \\
504010010 (Rim6) & 5.0 ($\pm$ 1.12) \\
500020010 (NE1) & 5.0 ($\pm$ 0.77) \\
500021010 (NE2) & 3.0 ($\pm$ 0.58) \\
500022010 (NE3) & 6.0 ($\pm$ 0.35) \\
500023010 (NE4) & 6.0 ($\pm$ 0.34) \\
\hline
\end{tabular}
%\end{minipage}
%\end{table*}
\end{table}

\section{Comparison with non-radiative H$\alpha$}\label{s:ha}

While radiative H$\alpha$ emission is associated with radiative cooling, non-radiative H$\alpha$ is associated with relatively faster shocks (v$_s\geq200$km/s) that heats the largely neutral upstream gas to a temperature that emits X-ray emission.  Such shocks are characterized by the lack of emission from forbidden lines of lowly-ionized metals. This non-radiative H$\alpha$ emission arises primarily from hot protons after electron capture via CX with neutral atoms and from collisional excitation of cold neutral hydrogen in the immediate post-shock region \citep{Levenson1998}.  With spectroscopy these two contributions can be separated due to velocity distribution differences of the two sources with CX producing a broad line and collisional excitation producing a narrow line.  However, both components are related to the interaction of neutrals with the hot plasma behind the shock front.  Therefore, the K$\alpha$ emission of OVII via CX should be correlated with regions of enhanced H$\alpha$ emission relative to low ionization metal forbidden lines.

\begin{figure}
\centerline{\epsfig{figure=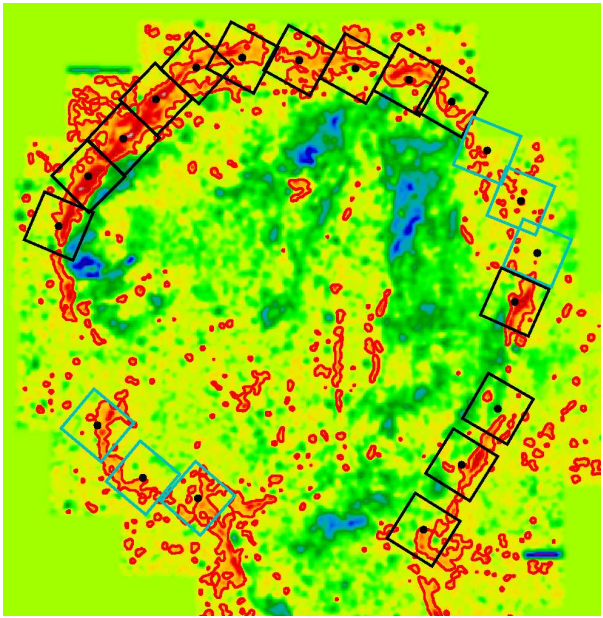,width=0.45\textwidth,angle=0}}
\caption{%\footnotesize 
Logarithmically scaled H$\alpha$/[S II], $\Re$, ratio map smoothed with a 3 pixel kernel.  Green is equivalent to a $\Re=1$, with red being the regions of large $\Re$ (relatively high H$\alpha$) and blue being the regions of small $\Re$.  The red contour shows the minimum ratio value to be considered non-radiative H$\alpha$, which we see are predominantly confined to the rim of the SNR.  Data courtesy Dr. Nancy Levenson (Levenson et al. 1998). }
\label{fig:Halpha}
\end{figure}

In order to compare with non-radiative H$\alpha$ emission of the Cygnus Loop SNR, we use the generously provided data presented in \cite{Levenson1998}.  However, the H$\alpha$ emission can arise from both radiative (cooling) and non-radiative regions.  To extract the structure and emission that is predominantly non-radiative, we use the ratio of two narrow band images, H$\alpha$ to [SII] ($\lambda\lambda$6717,6731) ($\Re$; Figure ~\ref{fig:Halpha}).  Notice that the regions dominated by H$\alpha$ are confined predominantly to the rim of the SNR. Since this emission is relatively weak and only occurs in a narrow region behind the shock, similar to the CX X-ray emission, it only becomes apparent where the path length through the emission region is long.  Visual inspection indicates that there is a clear correspondence with the regions of the centroid shift and we see that strong non-radiative filaments are lacking in the southeast and northwest regions of the SNR.

To quantify this correlation, we extract the \suzaku\ observation regions individually from the unsmoothed global H$\alpha$ to [S II] ratio map.  Non-radiative H-alpha emission is then found by computing dendrograms of the image with minimum threshold $\Re_{min}$=1.1 and a minimum number of pixels of 200, which was chosen by eye to only select true regions of filaments.  The algorithm selects out continuous structure starting from the pixel maxima with the given threshold criteria.  This selection criteria is highlighted by a red contour in Figure ~\ref{fig:Halpha}.

\begin{figure}
\centerline{\epsfig{figure=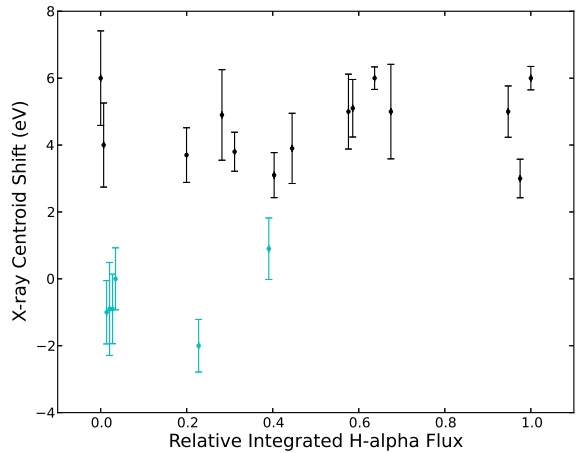,width=0.5\textwidth,angle=0}}
\caption{%\footnotesize 
OVII K$\alpha$ centroid shift between inner and rim regions ($E_{inner}-E_{rim}$) as a function of integrated H$\alpha$ flux density (normalized to 1) in observation region.  All data points of relative integrated H-alpha flux below 0.2 are in actuality at 0.  These points are staggered to allow one to see the different points.  We find that the presence of non-radiative H$\alpha$ filaments correlates with a centroid shift with 97\% certainty.  Reported error bars are 1 $\sigma$.}
\label{fig:shiftvhalpha}
\end{figure}

All structures found in the extracted regions are then combined.  These structures found via $\Re$ are then mapped to H$\alpha$ images and the integrated flux from the structures is calculated.  A comparison of this integrated flux with the centroid shift is shown in Figure ~\ref{fig:shiftvhalpha}. We see that observations without a centroid shift are confined to lower H$\alpha$ filament flux, with little filamentary structure.  Additionally, these X-ray centroid shift measurements appear to separate into an apparent dichotomy, those that have no shift and those that are forbidden line dominant and saturated.  While the determination of whether this saturation is physically reasonable is beyond the scope of this paper, we can ask the simple question of whether the presence of non-radiative H$\alpha$ filaments correlates with an OVII K$\alpha$ centroid shift, as we might expect if the soft X-ray emission is significantly contaminated by CX emission.

Binning the data based on whether there is an apparent centroid shift and on the presence of non-radiative H$\alpha$ filaments, we use Fisher's exact test to get a statistical handle on whether these two phenomena are correlated.  We calculate a p-value of 0.03, indicating significant likelihood of association between an OVII K$\alpha$ centroid shift and the presence non-radiative H$\alpha$ emission.  Note, the Rim5 observation was left out of the analysis due to a saturated point source, 52 Cygni, in the FoV of the unsmoothed optical image.

\section{Discussion}\label{s:dis}

\subsection{Comparison with Previous Work}

 Radially both this analysis and that of \cite{Katsuda2011} agree with the theoretical calcuations done by \cite{Lallement2004}, at least qualitatively.  In that work, the author shows that CX effects are generally negligible for an SNR in the Sedov phase, except in the narrow rim region where line of sight effects amplify the fraction of emission from CX.  In Figure 3 of that work, it is shown that CX can contribute over 50$\%$ of the X-ray emission immediately behind the shock and approximately 30$\%$ in the region contained within the shock radius(R$_{s}$) and r=0.98R$_{s}$ \citep{Katsuda2011}, thus detailing the confinement of CX to the rim.  However, azimuthally we find that the regions with OVII K$\alpha$ line centroid shifts do not match those with the emission enhancement at $\sim$ 0.7 keV, as studied by \cite{Katsuda2011}.  In fact, all of the regions that do not show an excess at $\sim$ 0.7 keV exhibit a centroid shift.

In overall flux, there is a great deal of tension between the CX interpretation of \cite{Katsuda2011} and the theoretical results of \cite{Wise1989}.  \cite{Katsuda2011} estimate the total CX contribution to the overall SNR emission to be 1-3 orders of magnitude higher than the prediction made by \cite{Wise1989}; the authors note, however, that the theoretical calculations were done with different shock velocities than those of the Cygnus Loop SNR.  There is also slight tension between our result and that of \cite{Wise1989}.  In that work they predict the CX contribution to the OVII K$\alpha$ line to be at most about 1\% of the line flux over the entire remnant.  If we assume that all of the line emission in the rim regions are due to CX and that the bulk of the emission of the SNR occurs in the outskirts of the SNR (within our observations shown), we roughly estimate the CX contribution to the \xs\ line to be $\approx$ 10\%, as an upper limit.

Beyond the reasonable agreement with theoretical work, our CX scenario is further confirmed by the correlation with non-radiative H$\alpha$ filaments.  Also, this analysis is far less model dependent.  It seems natural that this contamination of the X-ray spectra by CX may lead to the underestimate of the metal abundances when they are modelled with an oversimplified plasma model.  However, the lack of spatial correlation between the OVII K$\alpha$ shifts and 0.7 keV enhancement (which traces the abundance inhomogeneities) suggests that CX is not the only phenomenon missing in the modelling.  This is the same conclusion reached by \cite{Cumbee2014} where the authors attempted to do some modelling of CX emission in the remnant.  

\subsection{Resonance-line scattering}

There are a couple other mechanisms that could serve to increase the G ratio to the levels expected by CX. Resonance-line scattering is one such mechanism.  As seen in this work, the centroid energy shift is confined to the thin post-shock region.  This is also expected from resonance-line scattering as the optical depth may be greatly enhanced along tangential lines of sight through the immediate post-shock region where density is the greatest.  Resonance line scattering has been shown to be potentially important, causing the line intensity of the oxygen \ka\ triplet being underestimated by up to 40\% in regions that show apparently low metal abundance \citep{Miyata2008}. A reduction in the triplet intensity of this magnitude due to resonance-line scattering would result in an increase of the G-ratio to five assuming an initial G-ratio of one, as is typical of (nearly) thermal emission,  which would be more than enough to shift the line energies to the extent that we have measured.  

We examine the importance of the resonance scattering by confronting its expected and observed spatial and flux dependences.  In many of the regions that exhibit a centroid shift, the emission measure is larger in the inner spectrum than in the rim, while also having a temperature and ionization timescale that favors OVII as the dominant ionization state.  Consider a plasma of electron temperature equal to 0.21 keV, as is typical of our post-shock fits.  The main constituent throughout a wide range of ionization timescales would be OVII, from 1.$\times$10$^9$ s/cm$^3$ to 4.$\times$10$^{11}$ s/cm$^3$.  Once the plasma is equilibrated, the OVII fraction is still approximately 0.25.\footnote{These calculations were done using AtomDB 3.0, which is currently in a beta testing phase.  Its atomic physics is updated to include the effects of inner-shell ionization up to Nickel.}  Combining the mild change in OVII fraction with the larger interior emission measure, we would expect resonance line scattering to be more important interior to the rim. 

For example, in the NE1 region the best fit emission measures are 5.62$\times$ 10$^{57}$ cm$^{-3}$ in the rim ($EM_r$) and 1.06$\times$ 10$^{58}$ cm$^{-3}$ in the inner region ($EM_i$).  Assuming the regions have the same intrinsic oxygen abudance, the ratio of their optical depths can be written as:
\begin{equation}
\frac{\tau_r}{\tau_i}\approx\frac{f_{{\rm OVII},r}n_{e,r}L_r}{f_{{\rm OVII},i}n_{e,i}L_i}=\frac{f_{{\rm OVII},r}EM_r^{1/2}L_r^{1/2}}{f_{{\rm OVII},i}EM_i^{1/2}L_i^{1/2}}
\end{equation}
where f$_{{\rm OVII}}$ is the ionic fraction of OVII, n$_e$ the electron density, and L is the path length through the plasma.  Assuming a spherical remnant and using the path lengths at the inner sides of the extraction regions ($\sim$ 2' and 6') results in L$_r$/L$_i$ $\sim$ 0.53.  Using a conservative f$_{{\rm OVII},r}$/f$_{{\rm OVII},i}$ $\sim$ 2, equivalent to fully equilibrated in the inner spectrum and typical fractions of OVII for our post-shock ionization timescale fits, results in a strict upper limit of $\tau_r/\tau_i$ $\sim$ 1.1.

There are a couple reasons why this opacity ratio is likely to be an overestimate.  First, if one fixes the metal abundances in the rim to the local ISM values, the best fit emission measure is reduced by factor of 10, effectively reducing the calculated optical depth by a factor of 3.  Second, adding any missing line emission physics (e.g. CX) will likely serve to suppress the emission measure further, particularly in the rim since we expect this region to be further from equilibrium and thus more likely to suffer from missing physics.  Taken in conjunction, we expect the optical depth in the rim to actually be quite a bit lower than interior.  Even in the most favorable case, the optical depths in the rim and inner regions are comparable and we do not expect the differential resonance scattering between regions could result in our measured line centroid shifts.

\subsection{Inner-shell ionization}

\begin{figure}
\centerline{\epsfig{figure=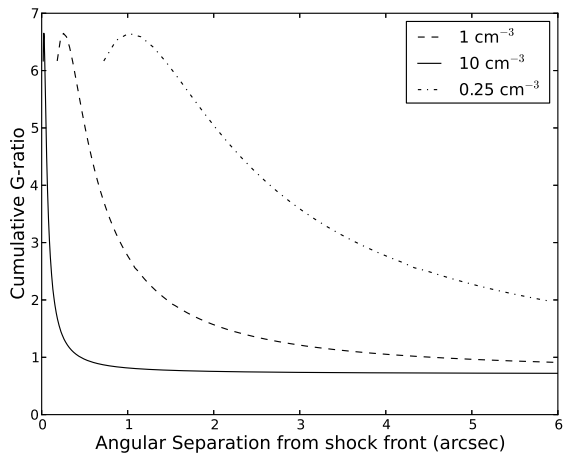,width=0.5\textwidth,angle=0}}
\caption{%\footnotesize 
The cumulative G-ratio plotted as a function of the angular width extending from the shock front when inner shell ionization of Li-like oxygen is included. Different lines correspond to different pre-shock gas densities, as specified in the insert, but assume the same characteristic shock velocity of 330 km/s and distance (540 pc) of the Cygnus Loop SNR.  This shock velocity is calculated from the typical electron temperature found in the immediate post-shock region (0.21 keV).
}
\label{fig:nei}
\end{figure}

Another process, inner shell ionization, is characteristic of an NEI plasma and has been shown to be important in the spectral modelling of young SNRs \citep{Kosenko2006}.  Through this mechanism, the G-ratio can be increased when the plasma is underionized and OVI is the main constituent.  When an OVI ion undergoes inner shell ionization an electron is ejected from the ground state, leaving one electron in the 1s $^{1}$S$_{0}$ state and the other in the 1s2s $^{1}$S$_{0}$ or 1s2s $^{3}$S$_{1}$ state.  This would typically lead to the forbidden transition and hence the G-ratio will be enhanced.  Figure ~\ref{fig:nei} demonstrates the potential effects of this process.

In this figure we calculate the cumulative G-ratio for a slit aperture of increasing width from the shock front toward the center of the SNR for a spherical shock of the size of the Cygnus Loop SNR.  The calculation is done spanning the range of surrounding ISM densities typically cited using updated atomic data provided by Borkowski\footnote{http://space.mit.edu/home/dd/Borkowski/APEC\_nei\_README.txt} to include inner-shell ionization and assuming infinite spatial resolution.  Note that the calculation also includes contributions from lithium-like satellite lines ($\approx$560eV), which are blended with the forbidden line.

Lower ISM densities push the curve to the right by lowering the ionization timescale at fixed distance from the shock front.  The G-ratio quickly peaks at $\approx$ 6.7, but decreases sharply beyond the peak and asymptotically approaches the collisional equilibrium value within several arcseconds, even for the favorable conditions considered here.  The state in which this effect is important occurs very early on, at n$_e$t $\approx 10^8$ s/cm$^3$, making this effect negligible for the angular resolution of the study presented here and unable to explain the \xs\ centroid shift observed.

\section{Conclusions and Future Prospects}

Plasma interfaces are regions rich in interesting physics.  We are just now straining to visualize their intricacies as we push the limits of our observational ability.  For instance, without being able to identify spectroscopic signatures of CX and inner-shell ionization, such as \ka\ G-ratios, the overall contributing effects have been elusive for X-ray astrophysics and it becomes difficult to properly infer the plasma properties, such as temperature and metal abundances, particularly for the CX case since it has yet to be well modelled.  Ideally, one wants to resolve the \ka\ triplets and hence to use its full diagnostic power.  Nevertheless, we can still utilize unresolved \ka\ triplets in cases of either moderate resolution spectra or low S/N high-resolution spectra to probe the physical conditions of plasma via their centroid, as demonstrated in the presented analysis of the Cygnus Loop SNR.   The conclusions of this analysis are as follows:
\begin{itemize}
\item The centroid shift between the on and off-shock OVII K$\alpha$ triplet can be naturally explained by CX in terms of its magnitude ($\approx 5$ eV) and radial distribution.
\item  The azimuthal correlation of the shift with non-radiative H$\alpha$ emission reinforces this interpretation.
\item  CX appears to be the only viable mechanism to explain the centroid shift.
\item  CX as traced by the centroid shift does not spatially correlate with, and therefore can't be the sole source of, the abundance inhomogeneities.
\end{itemize}
These conclusions seem to be firm, demonstrating the potential of this novel approach.  Even as we move into the future with the next generation of X-ray telescopes, all with high-resolution spectroscopic capabilities, approaches similar to the one presented in this paper will remain useful.  Astro-H, for example, will have an X-ray calorimeter onboard, with an expected angular resolution better than 1.3 arcmin and expected spectral resolution of $\sim$ 7 eV.  While such an instrument will undoubtedly be revolutionary for our understanding of diffuse X-ray emission, even with this resolution the triplet cannot be fully resolved, making centroid energy diagnostics vital to the future of X-ray spectroscopy.  Unfortunately, the small FoV of Astro-H (similar to the angular resolution), will prevent the specific approach presented here from being overly useful.  However, one may reasonably use consecutive observations with little fear of a change in instrumental gain.  Furthermore, CX is not expected to influence the energy of the OVIII K$\alpha$ line.  This allows us to leverage this line as an instrumental gain reference and probe any OVII forbidden line enhancement for the case where these two unresolved lines are well measured.

\section*{Acknowledgments}
We would like to thank Konrad Dennerl for a constructive referee report, as well as Nancy Levenson for providing the narrow-band optical images and Kazik Borkowski for providing updated atomic data.  We would also like to thank David Henley and Christina C. Williams for their useful discussions.  This research has been supported by NASA via the grant NNX12AI48G and has made use of data obtained from the Suzaku satellite, a collaborative mission between the space agencies of Japan (JAXA) and the USA (NASA). This research has also made use of data and software provided by the High Energy Astrophysics Science Archive Research Center (HEASARC), which is a service of the Astrophysics Science Division at NASA/GSFC and the High Energy Astrophysics Division of the Smithsonian Astrophysical Observatory.

\label{lastpage}

\bibliographystyle{aa}
\bibliography{SNR_CX}

\end{document}